\documentstyle[preprint,aps,epsf]{revtex}
\begin{document}
\preprint{
  \parbox{2.0in}{%
       hep-ph/9606236 \\
       MSU-HEP-60605  \\
                      \\}
}
\title{Jet Algorithms and Top Quark Mass Measurement}
\author{Jon Pumplin}
\address{
Department of Physics \& Astronomy \\
Michigan State University \\
East Lansing, MI \\
Internet address:  pumplin@pa.msu.edu
}
\date{\today}
\maketitle
\begin{abstract}
Mass measurements of objects that decay into hadronic jets, such as
the top quark, are shown to be improved by using a variant of the $k_t$
jet algorithm in place of standard cone algorithms.  The possibility
and importance of better estimating the neutrino component in tagged
$b$ jets is demonstrated.  These techniques will also be useful in the
search for Higgs boson $\to b \bar b$.
\end{abstract}
\narrowtext

\section {Introduction}
\label{sec:intro}

It is often necessary to measure the mass of an object that decays
into hadronic jets.  An example of current importance is the decay
of the top quark $t \to b W$, where the $b$ quark materializes
as a jet and the $W$ boson decays either leptonically or into two
light-quark jets.  The accuracy with which the jets can be measured
governs the error in the top quark mass measurement, which is
crucial to the study of electroweak physics --- e.g., knowing
$m_t$ allows a logarithmic estimate of the Higgs boson mass in the
minimal
model.  Accurate measurement of the jet decays of the $W$ is also
valuable here because good $W$ mass resolution can reduce the
combinatoric and other backgrounds in the analysis.  Futhermore, the
hadronically decaying $W$ can provide an alternative measure of
$m_t$ based on the jet angles in the top rest
frame \cite{improving}:  these angles determine $m_t / m_W$ in
each event with errors that are largely independent from the errors
of the traditional measure, so the two methods can be averaged to
improve resolution.  At the same time, $t \bar t$ events offer
a sample of hadronic $W$ decays that can be compared against the
known $W$ mass to test the theoretical and experimental assumptions
underlying all jet spectroscopy.  This opportunity is unique because
hadronic $W$ decays are otherwise obscured by large QCD backgrounds
and triggering problems \cite{cernw,howtosee}.

A second important application of jet spectroscopy occurs in the
search for Higgs boson $\to b \bar b$.  A moderate improvement in
dijet mass resolution has been shown to extend the
range of possible discovery to
$m_{\rm Higgs} \simeq 80 - 100 \, {\rm GeV/c^2}$ in
Tevatron Run II \cite{higgs}.

The important sources of error in jet spectroscopy are
(1) QCD radiation and hadronization effects,
(2) jet definitions, and
(3) detector effects.
We will compare these sources of error quantitatively, using
Monte Carlo simulation events for which the true partonic momenta
are known, and we will study the degree to which the jet finding
algorithm can be improved.  There is an interplay between the
first two sources of error because acceptable jet algorithms
differ from one another at next-to-leading order in $\alpha_s$
and in the nonperturbative hadronization corrections they
require.  Previous top quark analyses \cite{top} have used
cone algorithms for jet definition \cite{sterman,snowmass}.
But I will show in this paper that a particular version of the
$k_\perp$ successive recombination algorithm \cite{durham,soper}
instead promises superior results.

The detector effects studied here are generic ones that arise from
the basic segmented calorimeter design of all contemporary detectors.
Particular attention is paid to the unseen neutrino component of $b$
jets, which is found to be significant and partially correctable.
Dealing with the additional foibles of each specific apparatus
must be left to the experimentalists.

\section {Simulation}
\label{sec:simulation}
Throughout this paper we investigate the experimentally favorable
single-lepton ($\ell = e$ or $\mu$) top quark channel
$p \bar p \to t \bar t X$ with $t \to W^+ \, b \to {\rm jjj} \,$
and $\bar t \to W^- \, \bar b \to \ell^- \, \bar \nu_\ell j\,$
or their charge conjugates  at the present Tevatron energy
$\sqrt{s} = 1.8 \, {\rm TeV}$.  The results also apply rather
directly to the six-jet channel where both $t$ and $\bar t$
decay hadronically.

Because of color confinement, the quarks from top decay show
themselves as jets of hadrons \cite{barlow}.
One must infer the momenta of the quarks from measurements
of the observed jets.  Because of the collinear and soft
singularities of QCD, a quark naturally shares its momentum with
accompanying gluons and/or $q \bar q$ pairs.  It is necessary
to include these as much as possible in order to capture the
momentum of the original quark.
Sometimes the QCD radiation is so hard as to produce an extra
separate
isolated jet.  In such events, reconstructing the mass of the
original state is generally hopeless because the number of
combinatoric possibilities resulting from the many possible
sources of extra radiation is so large.  In many events, however,
the effect of the QCD radiation is simply to broaden the
jets in the $(\eta,\phi)$ plane.

Some of this territory has been explored previously \cite{seymour}.
However, we use here a significantly improved simulation program
with an up-to-date estimate of the top quark mass, and make a fuller
study of the effect of different options and parameters in the jet
definitions.  Also, we include the step of making ``jet energy
corrections'' which has become standard experimental practice.

\subsection {Event generation and cuts}
\label{sec:eventgen}
Events were simulated using the
{\footnotesize HERWIG 5.8} \cite{herwig}
Monte Carlo event generator, which models both hard and soft QCD
effects.  {\footnotesize HERWIG} is known to agree well with jet
data from $e^+ e^-$ interactions at values of $Q^2$ comparable to
those that arise in top quark decay \cite{leptuning}.  It also
agrees well with next-to-leading order perturbative calculations of
the distributions in $p_\perp^t$, $\eta_t$, and $m_{t \bar t}$ for
$t \bar t$ production \cite{mangano}.
The default {\footnotesize HERWIG} parameters were used, but I have
checked that substituting parameters that have been tuned to fit jet
data from $e^+ e^-$ interactions \cite{leptuning} causes negligible
change.
{\footnotesize HERWIG} does not include decay correlations between
the $t$ and $\bar t$ \cite{willenbrock}, or the finite width of the
top; but these effects are probably not important for our purposes.

Using {\footnotesize HERWIG} for top production is not without risk
in view of discrepancies with perturbative calculations that appear
specifically for top quark production \cite{orr}.  I have
incorporated a ``bug fix'' recently circulated by the authors of
{\footnotesize HERWIG} \cite{webber}, which substantially increases
the amount of hard gluon radiation in top decays and
removes the strong discrepancy shown in Fig.~2 of Ref. \cite{orr}.

I assume $m_t = 175$ in the simulation.
To approximate standard experimental cuts, I restrict the discussion
to events in which the lepton from $W$ decay has transverse momentum
$p_\perp^\ell > 20$ and pseudorapidity $|\eta| < 2$, and its neutrino
has $p_\perp^\nu > 20$.  These cuts keep $73 \%$ of the
single-lepton $t \bar t$ events.  (Units with ${\rm GeV} = c = 1$ are
used throughout this paper.)

Fig.~1 shows the $p_\perp$ distribution for the two $b$ quarks and
the two quarks from $W$ decay.  Typical values are comparable to those
for which {\footnotesize HERWIG} has been tested and tuned using
data from LEP \cite{leptuning}.  I impose a cut requiring all four of
these partons to have $p_\perp > 20$.  This cut keeps $67 \%$ of the
events that pass the lepton cuts.  It is intended to simulate the
effect of a cut on the minimum $p_\perp$ of the four highest $p_\perp$
jets observed in each event.  The cut is made at the parton level in
this simulation so that the different jet algorithms are compared
fairly, by applying them to the same set of events.  The partonic cut
should be very similar to experimentally possible cuts on observed jet
$p_\perp$ --- at least for the events that contribute to the signal,
for which the four highest $p_\perp$ jets in fact correspond to the
primary partons.

The reduction in signal due to a fairly strong cut on the minimum
$p_\perp$ of the observed four primary jets is a price worth
paying, particularly as the total number of observed events
rises, for several reasons:
(1) It avoids the need to measure jets of low $p_\perp$, which
have intrinsically large fractional uncertainties as is quantified
below;
(2) It increases the fraction of events for which the observed jets
will be correctly matched to their original partons, especially
since only the four jets with highest $p_\perp$ observed in each
event will be analyzed to reduce the combinatoric background in
assigning the jets; and
(3) $p_\perp$ cuts have been shown effective in suppressing the
major background from $W + {\rm jets}$ processes without
$t \bar t$ \cite{agrawal}.

\subsection {Detector models}
\label{sec:detectormod}
The detector is modelled as an array of $0.1 \times 0.1$ cells in
pseudorapidity $\eta = - \ln \tan \theta/2$ and azimuthal angle
$\phi$.  This granularity in the $(\eta,\phi)$ plane is similar to
that of the current {D\O} detector, while CDF detector cells have
width $0.26$ in $\phi$.
The detector is assumed to have no ability to identify particles,
so the energy deposited in each cell according to the simulation is
analyzed as if it came from a massless particle whose momentum
direction pointed toward the center of the cell.  (In real life,
corrections must be made for the spreading of energy into
neighboring cells due to the finite size of the shower generated
by a single particle.  This spreading also creates a possibility
in principle to locate the direction of momentum more accurately
than the cell size would predict.)

We consider three different models for the energy resolution of
the detector cells.  In model A (Ideal), the total energy deposited
in each cell is measured exactly, even including the
contribution from neutrinos.  In models B and C, the total energy
in each cell is smeared by realistic gaussian errors of standard
deviation $\Delta E$ given by
\begin{eqnarray}
\frac{\Delta E}{E} = \sqrt{\frac{c_1^{\,2}}{E} + c_2^{\,2}}
\label{eq:energyres}
\end{eqnarray}
with $c_1 = 0.55$, $c_2 = 0.03$ for charged hadrons (mostly $\pi^\pm$)
and  $c_1 = 0.15$, $c_2 = 0.003$ for $\gamma$, $e$ or $\mu$
(mostly $\gamma$ from $\pi^0$).  These parameters are approximately
those of the {D\O} detector \cite{D0nim}.

Models B and C differ only in that neutrinos are treated like
electrons in B, while in C the detector is blind to neutrinos like
a real detector.  The purpose for this distinction is that we will
find a sizeable difference between these two models because of the
frequent presence of neutrinos in $b$ jets, and it may be possible
to compensate for some of the neutrino component on an
event-by-event
basis using leptonic information that is acquired as a part
of some $b$ tags.

Cells that receive $p_\perp < 0.75$ are ignored in the analysis.
This mimics a limitation of the {D\O} detector due to noise levels
from its uranium calorimeter.  But it may be a good idea anyway
to drop contributions from very low $p_\perp$ particles, which
are at best poorly associated with any jet direction in
part because of hadron resonance decay effects and the difference
between rapidity and pseudorapidity; and because extraneous low
$p_\perp$ particles are present from soft hadronic interactions
that are additional to the hard
scattering that produced $t \bar t$ (``background event'')
and from independent $p \bar p$ interactions at high luminosity
(``pileup'').  The dependence on this $p_\perp$ threshold will
be discussed in Sect.\ \ref{sec:topmassres}.

Additional limitations that depend on experimental details of
real detectors, such as differences in the response to charged and
neutral particles in a shower, nonlinearity of that response, small
regions where there is no response, etc., are not included here.
The mass resolutions we find therefore represent an optimistic
limit for what can be expected.  However, the neglected effects
are generally small compared to those included, and they should
in particular not affect our conclusions on the relative merits
of different methods of analysis.

\subsection {Jet definition}
\label{sec:jetdef}
For jet spectroscopy, I advocate a particular version of the
$k_\perp$ jet finding algorithm \cite{durham,soper} that is
defined by the following explicit steps.
\begin{enumerate}

\item
Begin with a list of ``jets'' that consists simply of the
four-momentum from each cell above the $p_\perp > 0.75$ threshold,
treated as a zero-mass particle.  (There are typically
$\sim 40 - 60$ such cells, but more in a real detector where the
energy of a single particle is spread over several cells.)

\item
Compute $d_i$ for each jet and $d_{ij}$ for each pair of jets,
where $d_i$ is the jet transverse momentum and
\begin{eqnarray}
d_{ij} = \min(d_i,d_j) \, \Delta R / R_0
\label{eq:dij}
\end{eqnarray}
where
\begin{eqnarray}
\Delta R = \sqrt{(\eta_i - \eta_j)^2 + (\phi_i - \phi_j)^2}
\label{eq:delr}
\end{eqnarray}
is the angular separation in the $(\eta, \phi)$ ``Lego'' plane.
The parameter $R_0$ was introduced in Ref.~\cite{soper} to
generalize the $k_\perp$ algorithm.  It sets the scale for
the size of the jets in the $(\eta, \phi)$ plane.  Although it
does not create a sharp cutoff, cells that are farther than
$R_0$ from their final jet axis seldom contribute.
In this analysis, I mainly use $R_0 = 1$, which corresponds
to the original algorithm.  The dependence on $R_0$ will be
discussed in Sect.\ \ref{sec:topmassres}.

\item
Find the minimum of all $\{d_i,d_{ij}\}$.  If the minimum value
is less than $P_\perp^0$, the procedure is finished and the
current list contains the final jet momenta.  This termination
rule is different from some other versions of the $k_\perp$
algorithm.  The parameter $P_\perp^0$ defines a hardness scale
at which the algorithm terminates.  In particular, the final
jet list will contain no jets with $p_\perp$ below $P_\perp^0$.
I find that $P_\perp^0 = 10 \, {\rm GeV/c}$ works well for the
top quark analysis.

\item
Otherwise, if the minimum is a $d_i$, that jet is deemed to be
a fragment of one of the original beam particles (initial state
radiation) and it is dropped from the list.

\item
Otherwise, the minimum is a $d_{ij}$.  That pair of jets is
combined into a single jet by adding their four-momenta.

(The simple choice of adding the four-momenta to combine protojets
has an obvious good feature that the invariant mass of a multi-jet
object will be stable with respect to changing the assignment of a
cell or group of cells from one jet to another within the object.
A customary alternative to this choice is to combine protojets
according to the ``Snowmass Accord'' \cite{snowmass} formulae
\begin{eqnarray}
p_\perp &=& p_\perp^i + p_\perp^j \label{eq:snowm1}
\\
\eta &=& (\eta_i \, p_\perp^i + \eta_j \, p_\perp^j)
/ ( p_\perp^i + p_\perp^j) \label{eq:snowm2} \\
\phi &=& (\phi_i \, p_\perp^i + \phi_j \, p_\perp^j)
/ ( p_\perp^i + p_\perp^j) \label{eq:snowm3}
\end{eqnarray}
where $\phi_j$ must be shifted by $\pm \, 2 \pi$ here and in
Eq.~(\ref{eq:delr}) if possible to minimize $|\phi_i - \phi_j|$.
I find this rule to give slightly poorer mass resolution than
simply adding the four-momenta.)

\item
Go to step 2.

\end{enumerate}

Only the four highest $p_\perp$ jets found by the $k_\perp$
algorithm are used in the analysis.  This causes a very small
fraction ($\sim \! 2\%$) of events to be dropped immediately
because fewer than 4 jets are found.  This can happen even though
we are looking for jets down to $p_\perp = 10$ from partons with
$p_\perp > 20$, because one jet can split into two or more by
hard radiation, or because two jets can lie so close together in
$(\eta, \phi)$ that they appear as one.  (It will eventually be
desirable to keep more than the four highest $p_\perp$ jets, to
allow for initial state radiation at higher $p_\perp$ than one
of the four primary decay partons or hard radiation from the
$t$, $\bar t$, $b$, or $\bar b$ \cite{barger}, in order to test
our understanding of QCD radiation; but because of its
combinatoric richness, this will not be helpful for the mass
measurement.)

The four hardest jets are matched to the four original parton
momenta, which are of course known in the simulation, by trying
all $4! = 24$ assignments and keeping the one with the smallest
root mean square error in fitting the 4 parton directions in the
$(\eta, \phi)$ plane.  The jet energies are not considered in
this matching process, so as not to bias our study of the
accuracy of jet energy measurement.

The distribution in the rms error of the best fitting assignment
shows a strong peak at small values, above a background that extends
to large ones.  We impose a cut $\alt 0.8$ on the total rms error,
which is equivalent to a cutoff at $\alt 0.4$ for the average
deviation in $(\eta, \phi)$ from each of the four parton directions.
This cut keeps $67\%$ of the events.  The events it removes are
mainly those in which the four highest $p_\perp$ jets are not the
right ones because of initial state radiation of a gluon with higher
$p_\perp$ than one of the top decay quarks.  Thus our procedure of
keeping only the four jets with highest $p_\perp$ captures the desired
two $b$ jets and two $W$ decay jets about 2/3 of the time.

The events that survive the rms fit cut are used to study the
$p_\perp$ resolution for jets, and the resulting mass resolution
for $t \to bW \to {\rm jjj}$, in the next two sections.
To compare the effects of different jet algorithm parameters or
detector parameters fairly, the location of the cut is adjusted
slightly to keep the fraction of events that pass the cuts constant.

\subsection {Jet energy resolution}
\label{sec:jetenergyres}
Figs.~2--4 show the ratio
$p_\perp^{\rm \, Jet} / p_\perp^{\rm \, Parton}$ at
$p_\perp^{\rm \, Parton} \simeq 50$.  The solid curves are for
jets from $W$ decay (light quarks), while the dotted curves are
for $b$ jets.  The three Figures correspond to the three models
for calorimeter energy resolution:  Fig.~2 assumes perfect resolution,
while Fig.~3 and Fig.~4 both include the realistic energy resolution
given in Eq.~(\ref{eq:energyres}).  The detector is assumed capable
of detecting neutrinos in Figs.~2 and 3, while it is blind to them in
Fig.~4.

All of the curves peak at
$p_\perp^{\rm \, Jet} / p_\perp^{\rm \, Parton}$ below $1$ because
of the assumed $p_\perp$ threshold of the cells and because
QCD radiation can cause a significant fraction of the jet energy to
appear at large angles where it is omitted by the jet algorithm.
The peaks in Fig.~3 are more than twice as wide as the peaks in
Fig.~2.  This indicates that the energy resolution of the
calorimeter cells is the major source of error in the jet energy
measurement:  e.g., if the QCD and calorimeter cell size
errors included in Fig.~2 and the resolution errors were
equal, the peak width would increase only by a factor $\sqrt{2}$
in going from Fig.~2 to Fig.~3.

Fig.~2 shows only a small difference between $b$ jets (dotted) and
the light quark jets from $W$ decay (solid).  The difference remains
small when energy resolution is included in Fig.~3.  In going from
Fig.~3 to Fig.~4, there is almost no change in the $W$ decay jets
(solid), as expected because there is not much neutrino component
in light quark jets.  But a dramatic difference appears between
Fig.~3 and Fig.~4 for the $b$ jets (dotted).
{\it The loss in $b$-jet resolution due to varying amounts of
missing neutrino energy is very significant.  It will therefore
be useful to investigate the possibility of correcting for the
neutrinos on a jet-by-jet basis, using information that is
acquired as a part of $b$-jet identification.}

To study the dependence on partonic $p_\perp$, we can characterize
peaks like those shown in Figs.~2--4 by the value of
$p_\perp^{\rm \, Jet} / p_\perp^{\rm \, Parton}$ corresponding
to the $50^{\rm th}$ percentile (median) of the distribution, and
the values corresponding to the $16^{\rm th}$ and $84^{\rm th}$
percentiles which define the middle $68\%$ of the probability
distribution.  These would be the $\pm \, 1 \sigma$ points if the
distributions were Gaussian.  The result is shown in Figs.\ 5--7,
expressed in terms of the difference
$p_\perp^{\rm \, Jet} - p_\perp^{\rm \, Parton}$ instead of the
ratio for convenience.

One sees that the $50^{\rm th}$ percentile curves in Figs.\ 5--7
can be reasonably well approximated by straight lines.  Those
straight line fits can be used to make average ``jet energy
corrections'' of a linear form
\begin{eqnarray}
p_\perp^{\rm \, Parton} \simeq A \, + \, B \,p_\perp^{\rm \, Jet}
\label{eq:corrections}
\end{eqnarray}
to better estimate the partonic energy from an observed
jet energy.  The appropriate parameters $A$ and $B$ are somewhat
different for $b$ jets and $W$-decay jets, and vary with the
parameters of the jet algorithm.

After average jet energy corrections have been made,
fluctuations from jet to jet remain due to different amounts of
QCD radiation falling outside the identified jet.  These
fluctuations contribute
to the energy resolution errors, and hence to the width of peaks in
multi-jet mass distributions.  The ``$\pm \, 1 \sigma$''
spread in $p_\perp^{\rm \, Jet} - p_\perp^{\rm \, Parton}$ is seen
in Figs.\ 5--7 to grow only slowly with $p_\perp^{\rm \, Parton}$,
so the {\it fractional} accuracy of the $p_\perp$ measurement
improves significantly with increasing  $p_\perp$.  The spread in
$p_\perp^{\rm \, Jet} - p_\perp^{\rm \, Parton}$ is larger for $b$
jets.  This is dramatically so in the case of the most realistic
detector model C, which admits the possibility of large energy
escape in the form of neutrinos.

\subsection {top quark mass resolution}
\label{sec:topmassres}

We concentrate on the mass measurement of the hadronically
decaying top, since it is a good example of ``jet spectroscopy''
in general, and since the treatment of the leptonically decaying
top is complicated by errors in the measurement of the neutrino
momentum.  (The transverse momentum of the neutrino is
inferred from missing $p_\perp$, which can be strongly affected
by detector imperfections and by the presence of neutrinos in the
$b$ or $c$ jets.  The longitudinal momentum of the neutrino is
subsequently obtained by assuming $m_{\ell \nu} = m_W$, which
acquires serious uncertainties from the error in $p_\perp^\nu$
and the finite $W$ width in addition to the two-fold ambiguity
in the sign of $\eta_\nu - \eta_\ell$.)

Three-jet mass distributions from $t \to bW \to jjj$
are shown in Fig.~8 for the three
models of calorimeter energy resolution.  In generating these
histograms, the best match to the four parton directions was
again used to infer the jet assignments.  But this time
the best-fitting assignment is plotted for every event, without a
cut on the quality of the fit.  This makes the simulation more
realistic, since it includes backgrounds of a type that will be
present in actual data analysis.  The jet assignments are needed
to know which three of the four jets come from the hadronic
top decay, and also because linear jet energy corrections are made
using Eq.~(\ref{eq:corrections}) with parameters $A$ and $B$ that
are slightly different for $b$ jets and light-quark jets according
to Figs.\ 5--7.

Thanks to the jet energy corrections, the peaks are centered very
close to the input value $m_t = 175$.  Their shapes are not
symmetrical, but are instead skewed toward low masses
since QCD radiation and loss due to neutrinos can substantially
reduce the observed energy of a jet, but cannot increase it.
The widths of these peaks can be measured by fitting the histograms
to a Gaussian
plus a linear background over the fairly narrow mass range
$160 < M_{\rm jjj} < 190$: this is useful for purposes of comparison,
even though the resulting fits are not statistically adequate at the
high statistics at which the histograms have been computed.  The
resulting gaussian peaks correspond to
standard deviations of $\Delta M = 4.0$, $7.3$, and $9.1$ for the
three models of resolution.  Fitting over a different mass range
results in somewhat different numbers, but leads to the same
qualitative conclusions.

The mass resolution for $m_t$ can be improved by replacing the
usual invariant mass estimate, which is based on the sum of the
4-momenta
of the three jets, by the average of that value and a mass estimate
based on the jet angles in the top rest frame \cite{improving}.
Three-jet mass distributions obtained using this average
variable are shown in Fig.~9.  These peaks are more symmetrical
than those of Fig.~8 because fluctuations in the jet angle part of
the mass measurement have no definite sign.  The peaks are
narrower in each case, with widths $\Delta M = 3.9$, $5.7$, and
$7.3$ for the three models of resolution.  This demonstrates the
value of the jet angle method.

The dependence on the assumed calorimeter cell threshold is not
large.  For example, raising the threshold from $p_\perp > 0.75$
to $p_\perp > 1.00$ increases the width of the mass peak by only
$\simeq 5 \%$ in the case of model B for the energy resolution.
Similarly, lowering the threshold to $p_\perp > 0.50$ narrows the
peak by $\simeq 5 \%$.  The actual effect would be even less than
that because the ``background event,'' which contributes random
noise at low $p_\perp$, has not been included in the simulation.

The dependence on the jet radius parameter $R_0$ of the $k_\perp$
algorithm is also not large.  The original choice $R_0 = 1$ is
found to be close to optimal.  Going to $R_0 = 0.8$ or $R_0 = 1.2$
results in mass peaks that are a few percent broader.

One might wonder if the $k_\perp$ algorithm could be improved in
some cases by revising its assignment of cells to jets according
to their proximity to the jet axes it finds.  To test this, the
following plausible modification was tried:  After completing
the work of the $k_\perp$ algorithm on each event, any cell above
the $p_\perp > 0.75$ threshold was reassigned to the nearest of the
four highest $p_\perp$ jet axes if the cell was within $0.7$
of that axis and
(1) it was previously assigned to a different jet whose axis is
farther away than this new one by a factor $> 1.2\,$, or
(2) it was previously not assigned to any jet.
This modification affected only $17\%$ of the events, almost
entirely through option (1).  It produced a small improvement
in energy resolution for that subset of events, but the improvement
was not large enough to make it worthwhile to ``second-guess'' the
$k_\perp$ algorithm in this way.

\section {Comparison with cone algorithms}
\label{sec:cones}
The analysis of jet data at hadron colliders has traditionally
been done using cone algorithms, in which a jet is defined as
the final particles within a circle of fixed radius $R$ in the
$(\eta,\phi)$ plane.  A typical cone size is $R = 0.7 \,$; but
smaller values like $0.4$ have been used for processes like
$t \bar t$ production, to improve the sensitivity to
configurations where partons lie close together in the
$(\eta,\phi)$ plane at the expense of increased errors in the
partonic momentum measurement due to fluctuations in the QCD
radiation lying outside the cone.

Cone algorithms are not at all straightforward to design, nor even
to describe, because of ambiguities in how to treat situations in
which jets overlap.  Overlap occurs to some degree whenever two jet
axes lie within $2 \, R$ of each other in $(\eta,\phi)$, which
happens in the majority of events of the type we are considering.

I have repeated the analysis of Section \ref{sec:simulation} with
the $k_\perp$ algorithm replaced by a cone
algorithm \cite{howtotell} that begins with clustering based on
equivalence classes \cite{youssef}.
I have also repeated the analysis using a version of the cone
algorithm by Seymour \cite{seymour}, which is patterned after
current practice.  A cone radius $R = 0.7 \,$ was used in both
cases.  The results achieved by these two cone algorithms, which
are alike in intent but very different in implementation, are
strikingly similar to each other.

Cone algorithms generally do not allow the final jet momenta to
lie within $R$ of each other.  This leads to a significant loss
of events in the top analysis, where the nearest pair of the four
primary partons lie within $0.7$ of each other in $20\%$ of the
events.  It shows up quickly on repeating the analysis of
Sect. \ref{sec:simulation}, in that $27\%$ of the events
for the algorithm of Ref.~\cite{howtotell}, or $32\%$ for the
algorithm of Ref.~\cite{seymour}, are rejected because fewer than
the required four jets with $p_\perp > 10$ are found, as compared
to only $< 2 \%$ for the $k_\perp$ algorithm.  Furthermore, the
distribution of errors in the best fit to the partonic angles is
broader for the cone algorithms than for $k_\perp$.

For events in which the necessary four jets are found, both cone
algorithms
perform almost as well as the $k_\perp$ one.  In particular, the
final $M_{\rm jjj}$ distributions are quite similar to those
shown in Figs.\ 8--9, especially for the cases in which realistic
calorimeter energy resolution is included, which masks the
differences.  The average energy corrections needed for the
cone algorithms are also similar to those for the $k_\perp$
algorithm, although slightly larger.

One could therefore say that the $k_\perp$ algorithm provides only
slightly better mass resolution than the cone algorithms, but allows
approximately $30 \%$ more events to be kept.  Another way to compare
the algorithms would be to impose a cut on the minimum separation
between observed jets in $(\eta,\phi)$ for the $k_\perp$ algorithm,
or to raise the $p_\perp$ threshold for defining jets in it, or to
make a combination of such cuts that would make the fraction of events
kept by the various algorithms the same.  The benefits of the $k_\perp$
algorithm would then appear entirely in the form of improved mass
resolution.

The solid curve in Fig.~10 shows the fraction of events for which
a good match is found between the 4 highest $p_\perp$ jets found by
the $k_\perp$ algorithm and the 4 primary partons (using a criterion
based on the quality of fit to the $(\eta,\phi)$ direction and
$p_\perp$ of all four) as a function of the minimum separation between
jets as observed by the algorithm.  The algorithm is seen to have
significant success even at minimum separations below $0.5$.
Meanwhile, the two versions of cone algorithm with $R = 0.7$
(dotted and dashed curves in Fig.~10) are
somewhat less effective overall, and are completely
unable to see separations smaller than the assumed cone size.
A smaller cone size could be used to extend the effectiveness of the
cone algorithms to smaller minimum separation, as CDF and D0 have
both done; but that would reduce the accuracy of the $p_\perp$
measurements, and hence reduce the overall fraction of good matches.
(As an aside, the curves shown in Fig.~10 are seen to turn over at
large minimum separation.  This may at first sight be puzzling, but
it only reflects the fact that large separation between all 6 pairs
of partons is very unlikely, so if the jet finder sees such a
configuration, it is likely to be mistaken.)

In setting up the definitive top quark data analysis, the best choice
of cuts on minimum jet-jet angular separation and minimum jet $p_\perp$
will have to be determined using a full simulation of both the detector
and the complete analysis procedure.  Optimal choices for the cuts for
the purpose of mass measurement will also depend on the number of
events available for analysis, since one can afford statistically
to cut harder when there are more events to begin with.

Another way to compare the $k_\perp$ and cone algorithms was carried
out to study the ability to analyze objects that decay into two jets
in the presence of additional jets, which will be necessary in the
Higgs boson search.  For this study, $t \bar t$ events were generated as
before except for an additional cut requiring the partons from $W$
decay to be separated from each other by $>1.0$ in the $(\eta, \phi)$
plane.  This cut is minor because these jets tend to be opposite
each other in azimuthal angle and hence well separated.
The ideal calorimeter model was used.  The events were
analyzed as before except that all jets found by the jet finder
were kept and there was no requirement that four or more jets be
found.  The pair of jets (at least two jets were always found) making
the best fit in $(\eta, \phi)$ to the two partons from $W$ decay
were identified.  Linear jet energy corrections were applied as
before to these jets.  The invariant mass of the pair was computed and
corrected for the deviation of the partonic $W$ mass from its nominal
value, to remove the effect of finite $W$ width that is included in the
simulation.  The resulting distribution in dijet mass is shown in
Fig.~11 for the $k_\perp$ algorithm and the two versions of cone
algorithm.  The distributions are normalized to the same number of
events, so the superiority of the $k_\perp$ method is demonstrated by
the fact that its peak is significantly higher.  This is true even
though the {\it width} of the peak ---  measured by full width
at half maximum above background or ``by eye'' --- is not obviously
better.  The point is that many events are so clean that all three
jet finders give almost identical results for them.  This can be seen
in Fig.~12, which shows the distribution of the total root-mean-square
deviation between the two jets identified as coming from the $W$ and
their true parton directions, i.e. the quantity that was
minimized to identify the ``correct'' jet pair.  Compared to the
$k_\perp$ algorithm, the two cone algorithms both have relatively
strong tails into a region of large deviation where the $W$ decay axes
have not been located very well.  These tails result mainly from events
in which the jet finder includes contributions to a $W$ decay jet from
particles actually coming from a $b$ jet that happens to lie nearby in
the $(\eta,\phi)$ plane.  This explains the tails extending toward
higher $M_{\rm jj}$ in Fig.~11.  The $k_\perp$ algorithm is less easily
confused by such particles.

\section {Neutrino momentum distributions}
\label{sec:neutrinos}
Figs.\ 8--9 show that there is a substantial loss in mass resolution
caused by fluctuations in the neutrino component of $b$ jets.  To
study this in more detail, Fig.~13 shows the distribution of the
observable (i.e., non-neutrino) fraction of jet momentum
\begin{eqnarray}
z = 1 \, - \, p_\perp^{\rm Neutrinos} / p_\perp^{\rm \, Parton}
\label{eq:zdef}
\end{eqnarray}
for $b$-jets that contain at least one neutrino.  The log-log plot
reveals that the distribution can be rather well approximated
by a power law:
$dP/dz \propto z^A$ with $A = 4.4$
for $z < 0.98\,$.  The dotted curve in Fig.~13 shows the
distribution for the subset of jets that contain an $e^\pm$ or
$\mu^\pm$ with $p_\perp > 2$, which may be detected
experimentally --- especially in the case of $\mu^\pm$.  The two
distributions are nearly identical.  Distributions with stronger
or weaker cuts on the $p_\perp$ of $e^\pm$ or $\mu^\pm$ , or with
cuts on $p_\perp^{\rm \, Parton}$, are also about the same.

We can use this power law over the entire range $0 < z < 1$ because
the neutrino contribution to $p_\perp$ is small compared to other
errors in jet energy measurement in the tiny region $0.98 < z < 1$
where the power law doesn't fit well.  Including the
contribution from jets without neutrinos then gives a normalized
parametrization of the distribution in observable momentum fraction
\begin{eqnarray}
\frac{dP}{dz} = f \, \delta(z-1) + (1-f) \, 5.4 \, z^{4.4}
\label{eq:powerlaw2}
\end{eqnarray}
where $f$ is the fraction of jets with negligible or zero neutrino
contribution.  For all $b$ jets, $f = 0.59$ which implies that
$23\%$ of them hide $> 10 \%$ of their momentum in
neutrinos and $12\%$ of them hide $> 20 \%$.
For the $33\%$ of $b$ jets that contain an electron or muon with
$p_\perp > 2$, $f$ is only $0.10$ which implies that
$51\%$ of them hide $> 10 \%$ of their momentum in
neutrinos and $27\%$ of them hide $> 20 \%$.
It is thus clearly advantageous to use different estimates to
correct for the missing neutrino energy in a $b$ jet, depending
on whether or not a lepton is observed in the jet.  This has already
been done in the analysis of the top quark signal \cite{bigcdf}.
A topic worthy of future study would be to see if any further details
of the observed jet, in addition to the mere presence or absence
of a lepton, can be used to further improve the neutrino momentum
estimate.

It is interesting that the distribution in missing neutrino energy
fraction
when a lepton is observed is nearly independent of the energy of
that lepton, except for the difference in probability that
the missing energy is negligible or zero.
Additionally, the probability distribution for the error in
jet momentum measurement is very asymmetric and very far from
gaussian.  This should be taken into account in the
$t \bar t$ final state reconstruction analysis.

\section {Direct Comparison of Mass Distributions}
\label{sec:direct}
So far, we have compared jet algorithms by making explicit use of
the original parton momenta to infer the correspondence between jets
and partons.  This facilitates a detailed comparison of the methods,
but it is somewhat artificial, since it can never be carried out
using real data for which underlying partonic information is unknown.
In this section we compare the jet algorithms directly, using no
information that exists only in the world of Monte Carlo.

An appealing way to make the comparison would be to simulate a full
data analysis recommended for $t \bar t$ events, and see how the
choice of jet algorithm affects the uncertainty in measuring $m_t$.
The treatment of measurement errors in that analysis, however, is
very complicated; and further complications arise from the role
played by missing $p_\perp$ in identifying the leptonically
decaying $W$, and from the existence of a variety of classes of
events with regard to $b$-tagging information (zero, one, or two
tags with varying degrees of certainty).  The complete comparison
can therefore only be done properly by the experimentalists who are
in a position to use full simulations of the detector, and who can
make the comparison with data as well as with Monte Carlo events.

In order to test our methods directly, but without carrying out the
full $t \bar t$ analysis, {\footnotesize HERWIG} events were
generated as before except that all parton-level cuts were removed.
The intermediate model of the calorimeter was used, i.e.,
energy resolution was included, but neutrinos were assumed to be
observable.  Instead of
using partonic information to infer the jet assignments, a trijet
mass distribution was found by simply plotting a histogram of
$M_{\rm jjj}$ formed from each subset of 3 of the 4 highest
$p_\perp$ jets.  Events with fewer than 4 jets were ignored.
The minimum jet $p_\perp$ was chosen slightly differently for
the different jet algorithms to make the fraction of events kept
the same for each algorithm.

The histograms of $M_{\rm jjj}$ are shown in Fig.~14.
The $k_\perp$ algorithm (solid curve) produces a clear peak above
the combinatoric background.  That background is very large because
even events that are analyzed correctly contribute three incorrect
combinations to the histogram in addition to the correct one.  The
two cone algorithms (dashed and dotted curves) produce nearly
identical results.  They show a peak that is significantly smaller
and broader than the result of the $k_\perp$ algorithm.

In a full analysis, $b$-tagging and the constraint from the hadronic
$W$ decay mass would greatly reduce the combinatoric background is
Fig.~14, and accentuate the difference between the methods.  The
signal peaks would also be slightly narrower because different jet
energy corrections could be made for the $b$ quark and light quark
jets, in place of the cruder method of just making an average
correction for all jets, which was used in generating Fig.~14.

\section {Conclusions}
\label{sec:conclusions}
We have seen that a form of the $k_\perp$ successive recombination
jet algorithm offers a significant improvement in the fraction of
$t \bar t$ events that can be reconstructed and/or offers
significantly improved
$t$ mass resolution at the same efficiency, compared with cone
algorithms like those that have been used up to now for $t \bar t$
data
analysis.  The basis of this is the flexibility of the $k_\perp$
algorithm with respect to jet radius:  it can include final
particles in a cone as large as $R = 1$ or even greater when
possible, while maintaining some useable efficiency for resolving
jets down to as close as $R=0.2\,$.
The improved mass resolution that can be obtained using jet angle
variables in the top rest frame \cite{improving} has also been
confirmed.
The size of these improvements and the importance of an accurate
top quark mass measurement are such that the procedure should be carried
out in spite of the considerable work that will be necessary to
reevaluate the instrumental corrections using the new methods.

The particular form of the $k_\perp$ algorithm advocated here
is characterized by a simple rule for when to terminate the process
of combining protojets into jets, as described explicitly in
Sect.\ \ref{sec:jetdef}.  The dependence on parameters appearing
in the algorithm is discussed in Sect.\ \ref{sec:topmassres}.  With
this algorithm, the mass resolution is close to optimal in the sense
that the majority of the width of the final mass peak is generated
by the nominal energy resolution of a typical detector, so not much
further improvement is theoretically possible.

We have seen that fluctuations in the momentum carried by neutrinos
contributes significantly to the error in measuring the momentum of
a $b$ jet.  This error is reduced in current practice \cite{bigcdf}
by using different distributions according to whether or not a
lepton is identified in the jet.  A matter for future study is
to see if any other features of the observed jet can be used to
further improve the estimate.

Finally, both the improved jet algorithm and the improved estimate
of neutrino contributions can help also in the search for other heavy
objects that decay into jets, such as
Higgs boson $\to b \bar b$ \cite{higgs}.

\section*{Acknowledgments}
I thank S. Kuhlmann, D. Chao, members of CTEQ, and members
of the {D\O} top quark mass group for discussions. This work was
supported in part by U.S. National Science Foundation grant
number PHY--9507683.


\newpage

\begin{figure}
    \begin{center}
       \leavevmode
       \epsfxsize=1.0\hsize
      \epsfbox{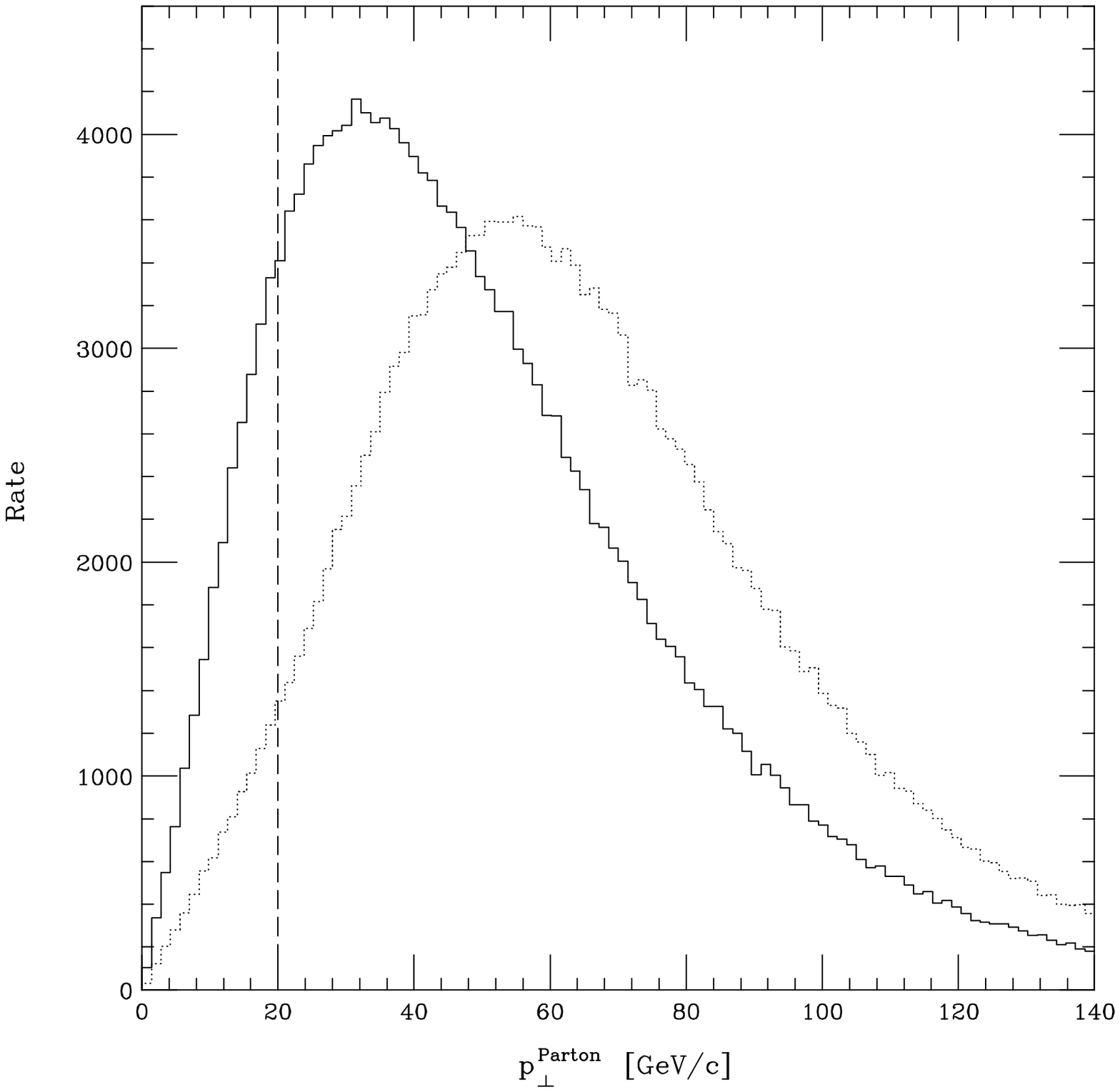}
    \end{center}
\caption{\protect
$p_\perp$ distributions in $t \to b W \to b q \bar q$ for
quarks from $W$ decay ({\it solid}) and $b$ quarks ({\it dotted}).
The dashed line shows the minimum $p_\perp$ cut used in this study.
}
\label{figure1}
\end{figure}

\begin{figure}
    \begin{center}
       \leavevmode
       \epsfxsize=1.0\hsize
      \epsfbox{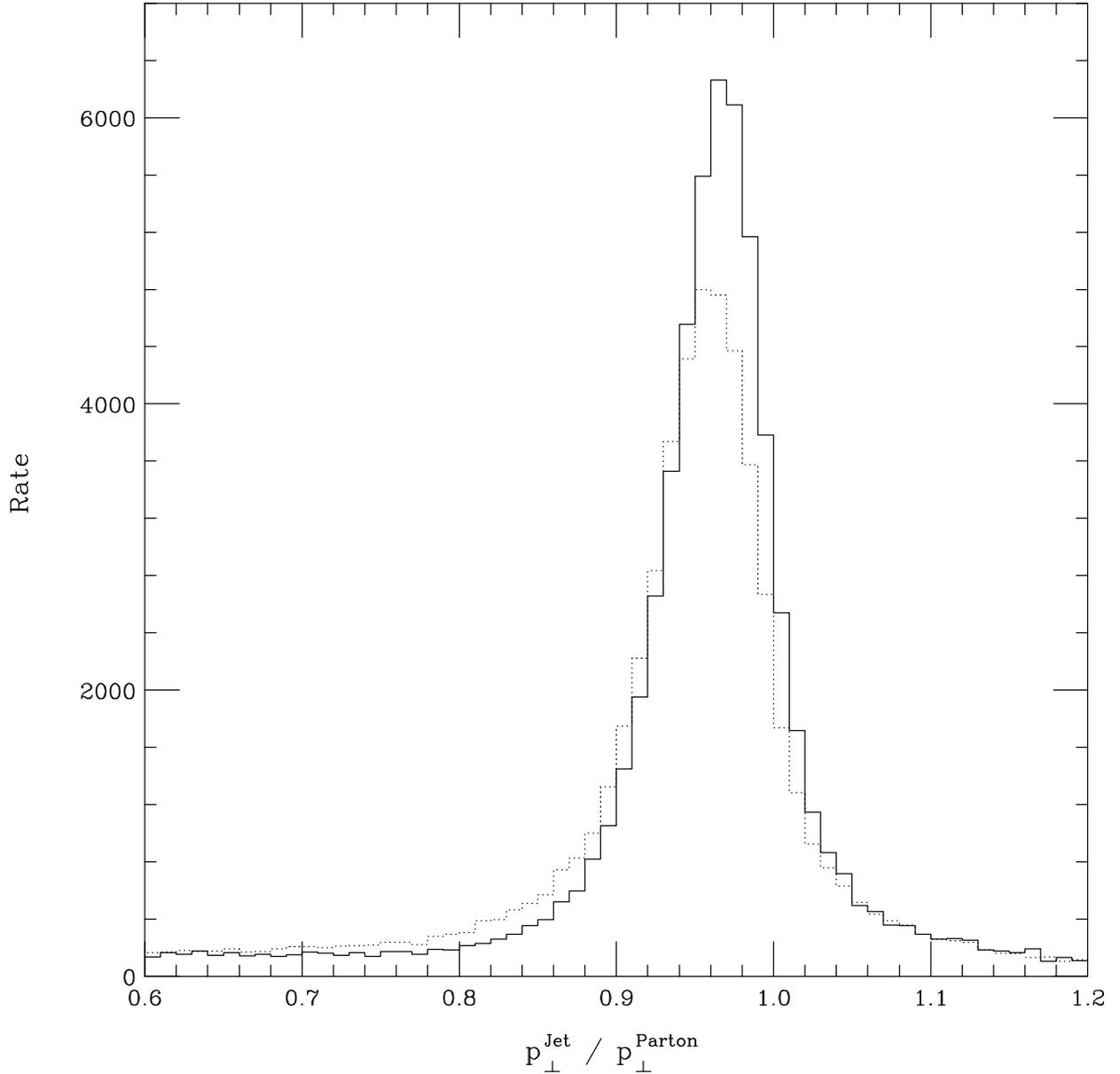}
   \end{center}
\caption{\protect
Distribution of the ratio of observed jet transverse momentum
($p_\perp^{\rm \, Jet}$) to original parton transverse momentum
($p_\perp^{\rm \, Parton}$) in $t \to b W \to b q \bar q$ for
quarks from $W$ decay ({\it solid}) and $b$ quarks ({\it dotted})
at $p_\perp^{\rm \, Parton} \simeq 50 \, {\rm GeV/c}$, for the
ideal calorimeter model.
}
\label{figure2}
\end{figure}

\begin{figure}
    \begin{center}
       \leavevmode
       \epsfxsize=1.0\hsize
      \epsfbox{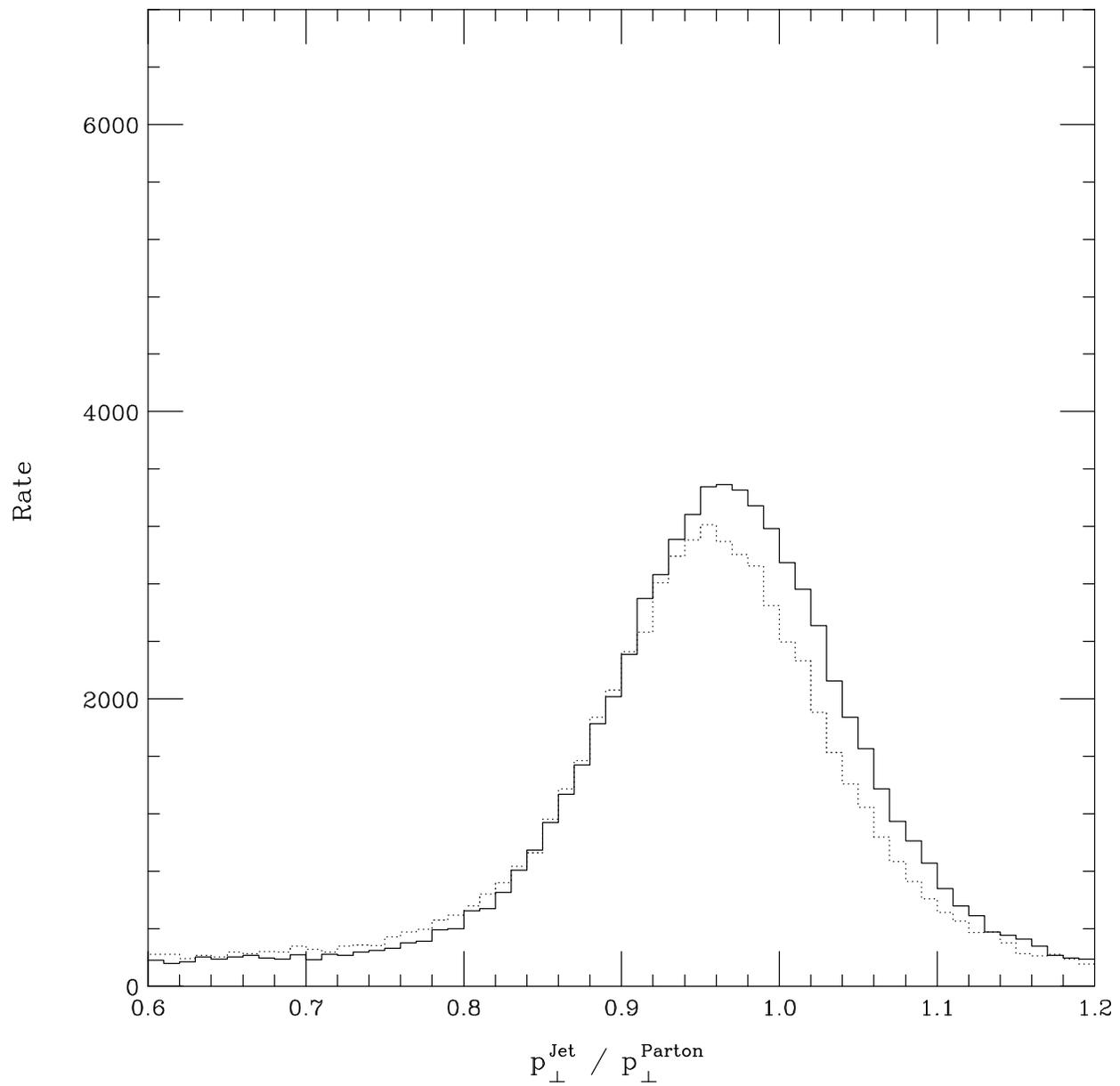}
   \end{center}
\caption{\protect
Like Fig.~2 except that the calorimeter model includes realistic
energy resolution.
}
\label{figure3}
\end{figure}

\begin{figure}
    \begin{center}
       \leavevmode
       \epsfxsize=1.0\hsize
      \epsfbox{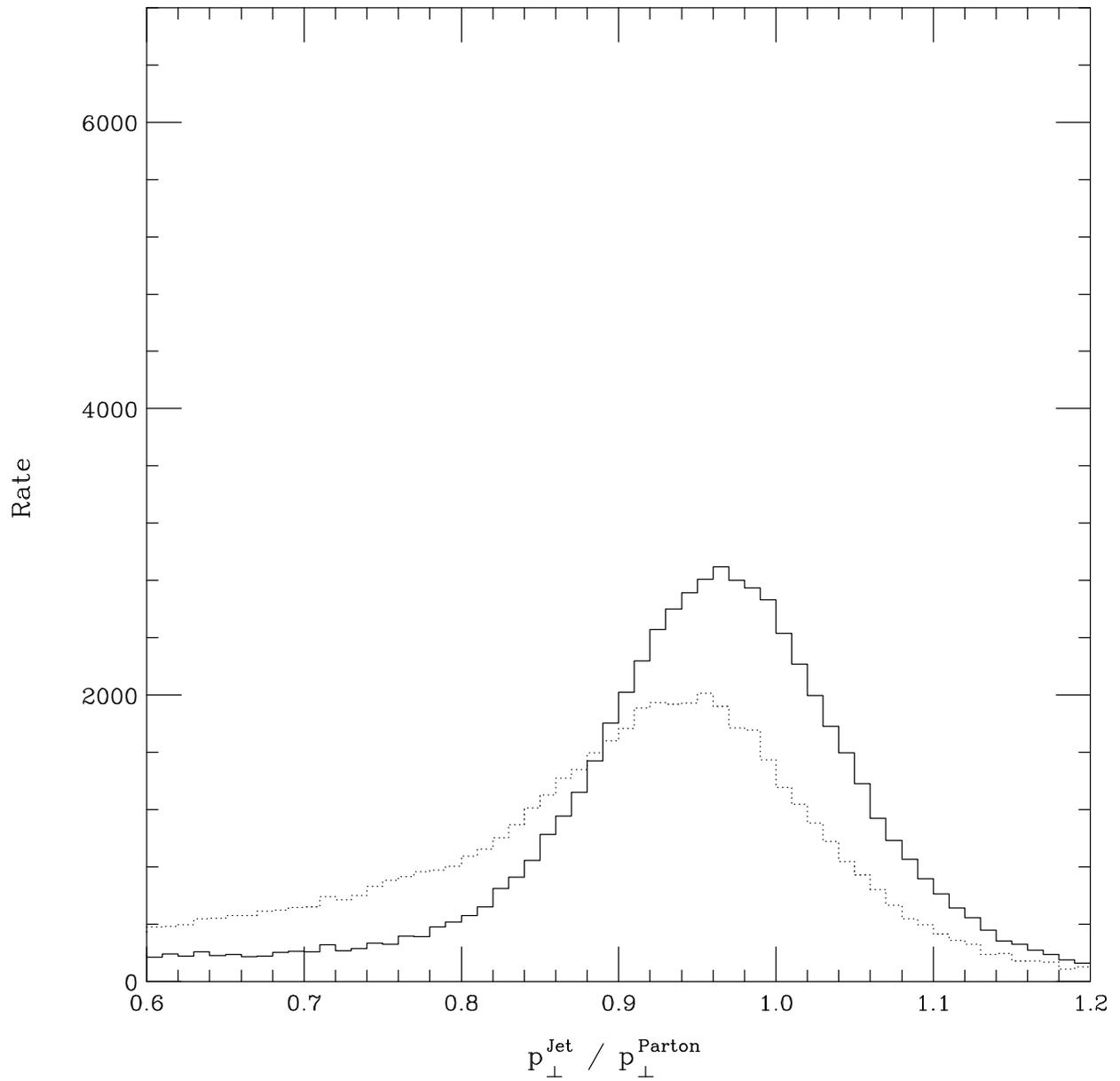}
   \end{center}
\caption{\protect
Like Fig.~3 except that the calorimeter is blind to neutrinos, which
is realistic unless the neutrino component can be estimated from
leptonic information.
}
\label{figure4}
\end{figure}

\begin{figure}
    \begin{center}
       \leavevmode
       \epsfxsize=1.0\hsize
      \epsfbox{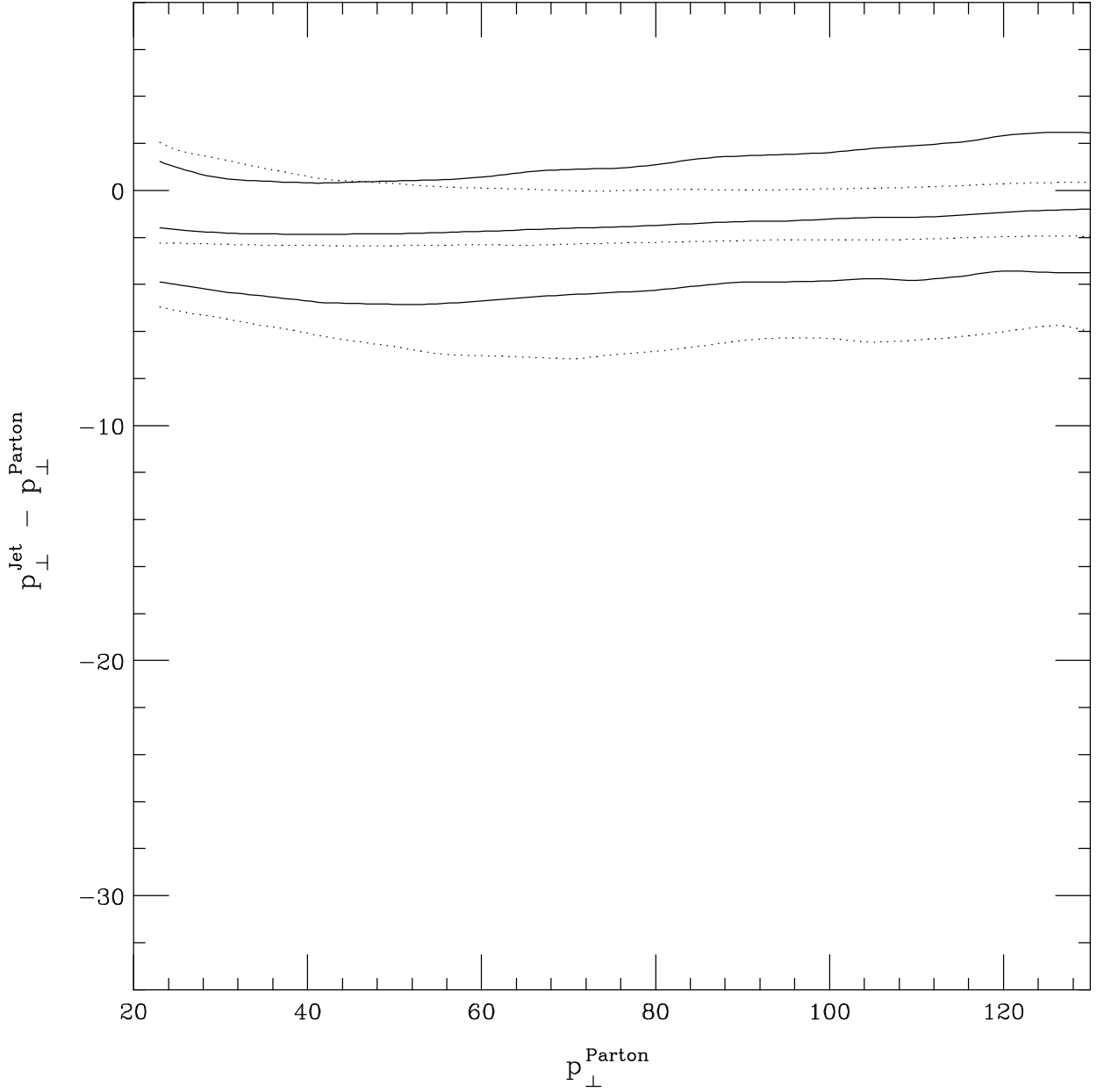}
    \end{center}
\caption{\protect
Three solid curves for $W$ decay jets and three dotted curves for
$b$ jets show the $16^{\rm th}$, $50^{\rm th}$, $84^{\rm th}$
percentile points (i.e., the middle $68\%$) for the distributions
of $p_\perp^{\rm \, Jet} - p_\perp^{\rm \, Parton}$ as a function
of $p_\perp^{\rm \, Parton} \,$.  The calorimeter model is the ideal
one as in Fig.~2.
}
\label{figure5}
\end{figure}

\begin{figure}
    \begin{center}
       \leavevmode
       \epsfxsize=1.0\hsize
      \epsfbox{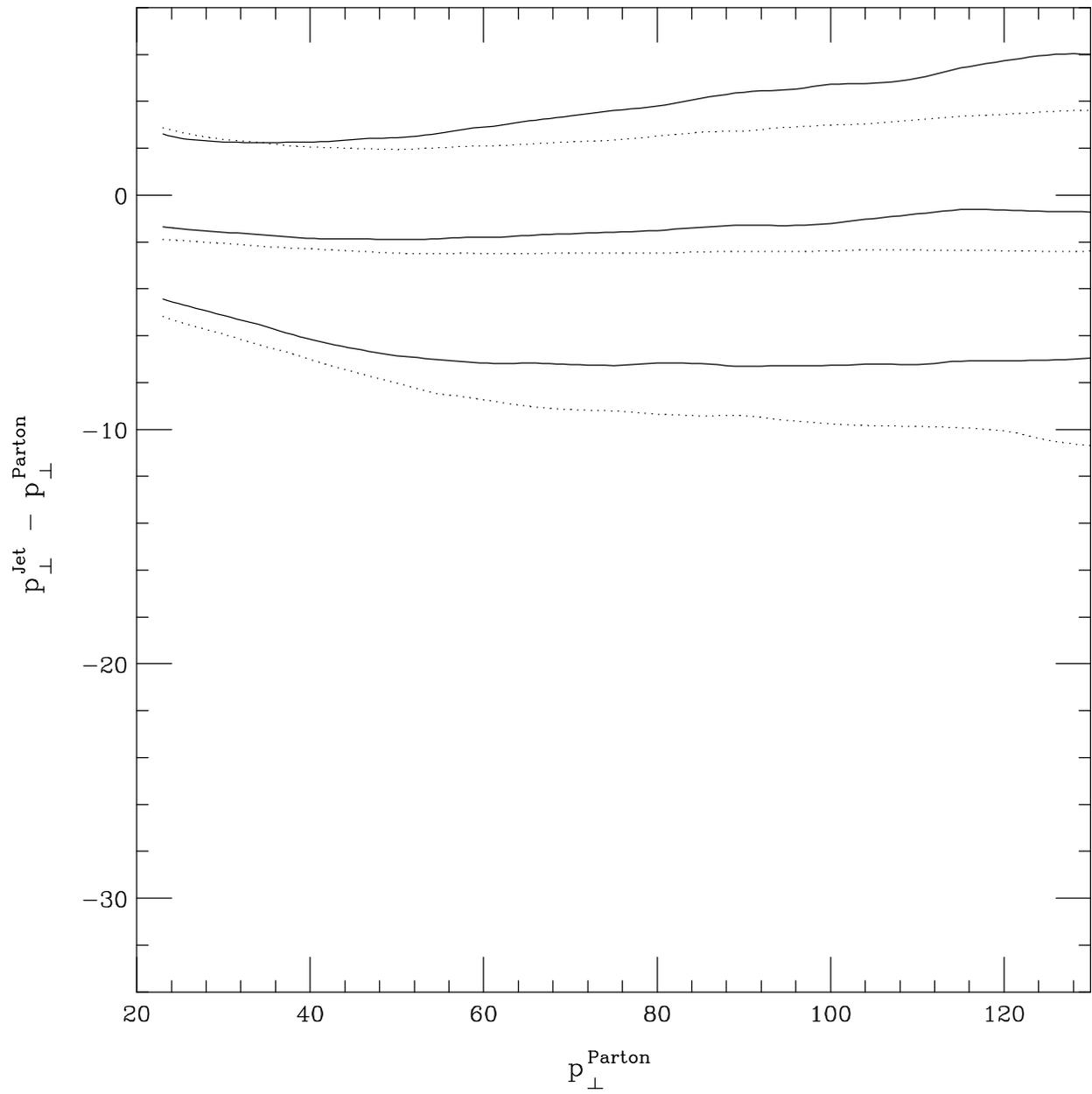}
    \end{center}
\caption{\protect
Like Fig.~5 but the calorimeter model includes energy resolution
as in Fig.~3.
}
\label{figure6}
\end{figure}

\begin{figure}
    \begin{center}
       \leavevmode
       \epsfxsize=1.0\hsize
      \epsfbox{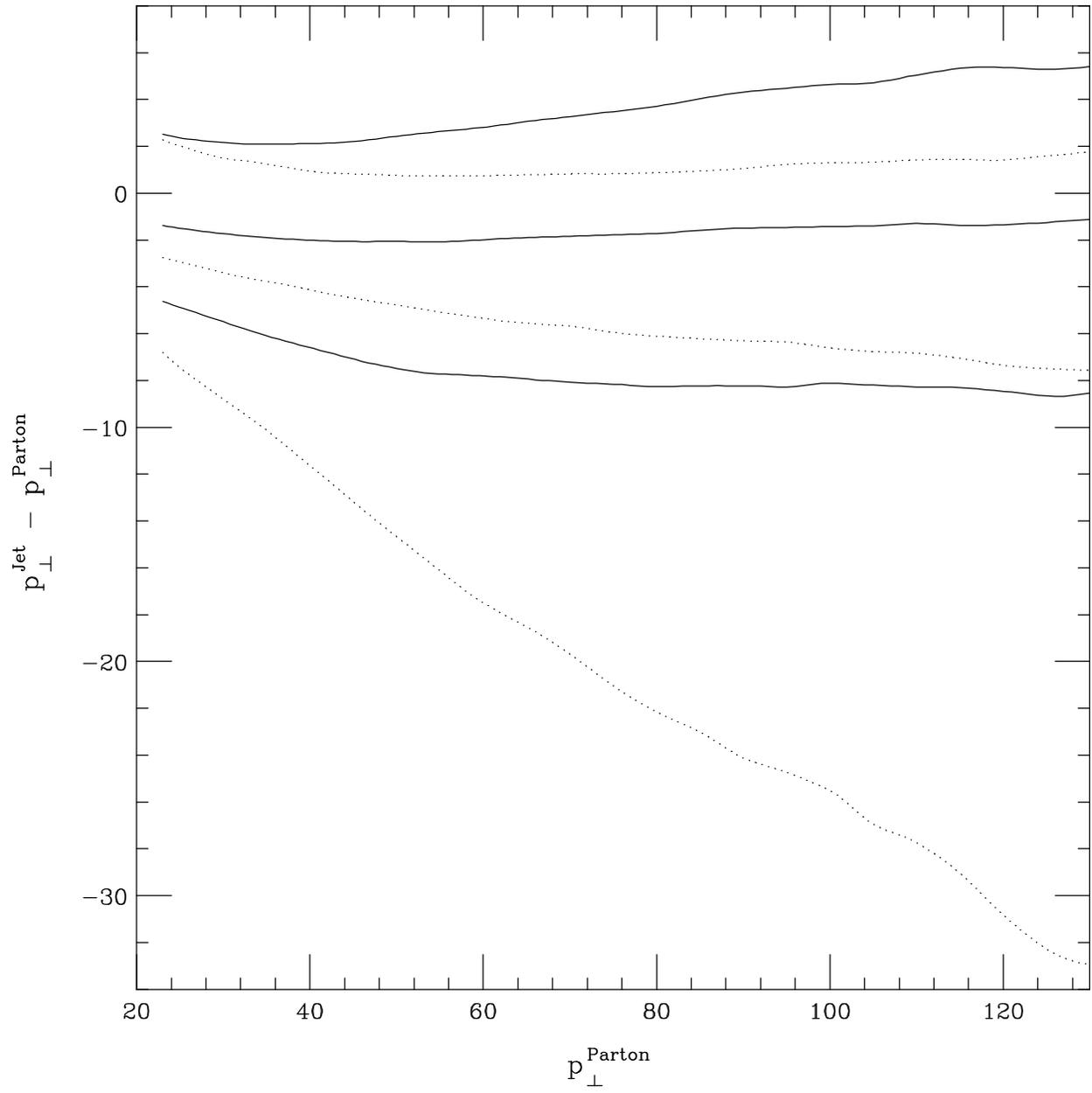}
    \end{center}
\caption{\protect
Like Fig.~6 but the calorimeter is blind to neutrinos as in Fig.~4.
}
\label{figure7}
\end{figure}

\begin{figure}
    \begin{center}
       \leavevmode
       \epsfxsize=1.0\hsize
      \epsfbox{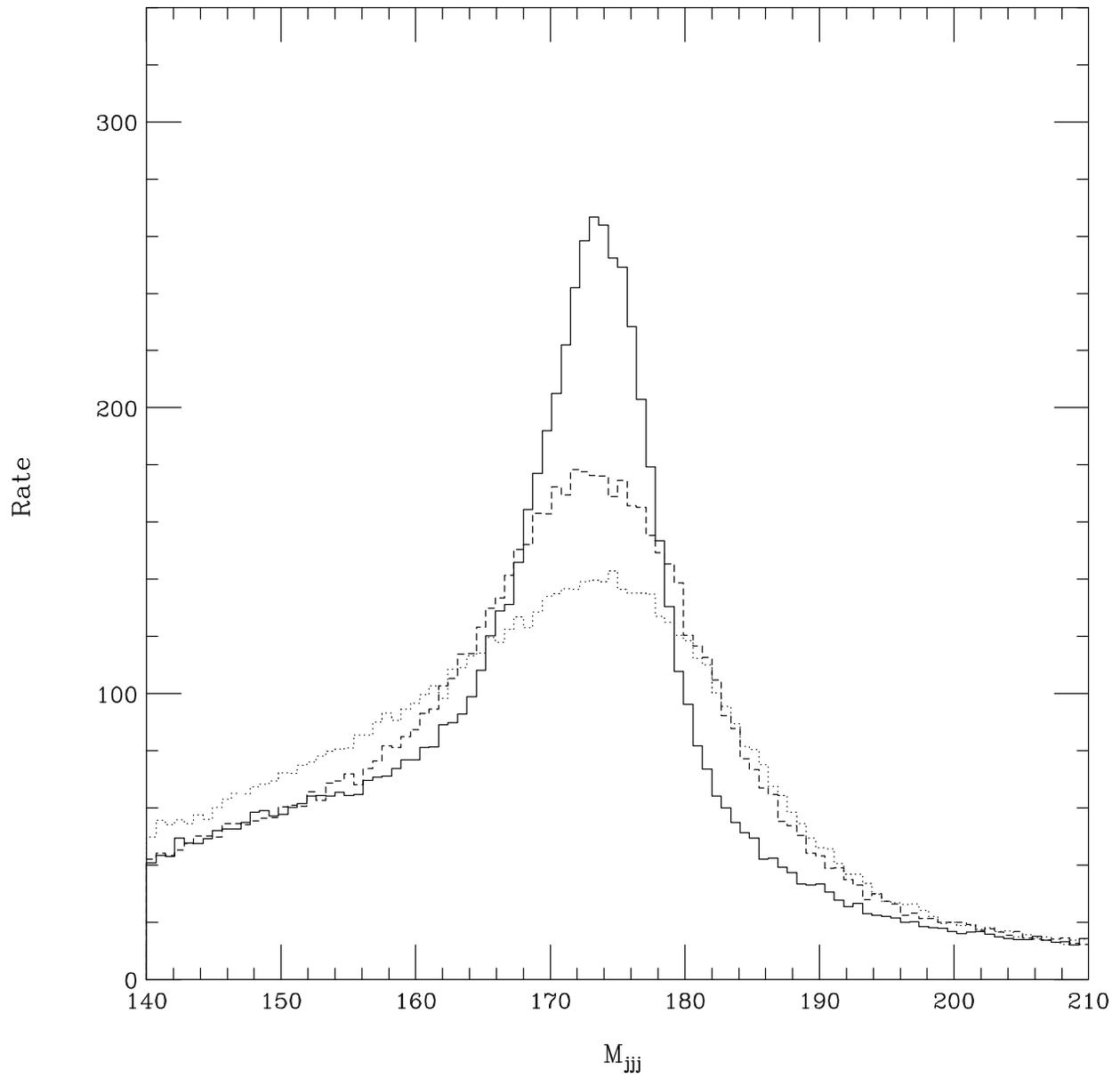}
    \end{center}
\caption{\protect
Invariant mass distribution for $t \to {\rm jjj}$ for the three
models of calorimeter energy resolution.
}
\label{figure8}
\end{figure}

\begin{figure}
    \begin{center}
       \leavevmode
       \epsfxsize=1.0\hsize
      \epsfbox{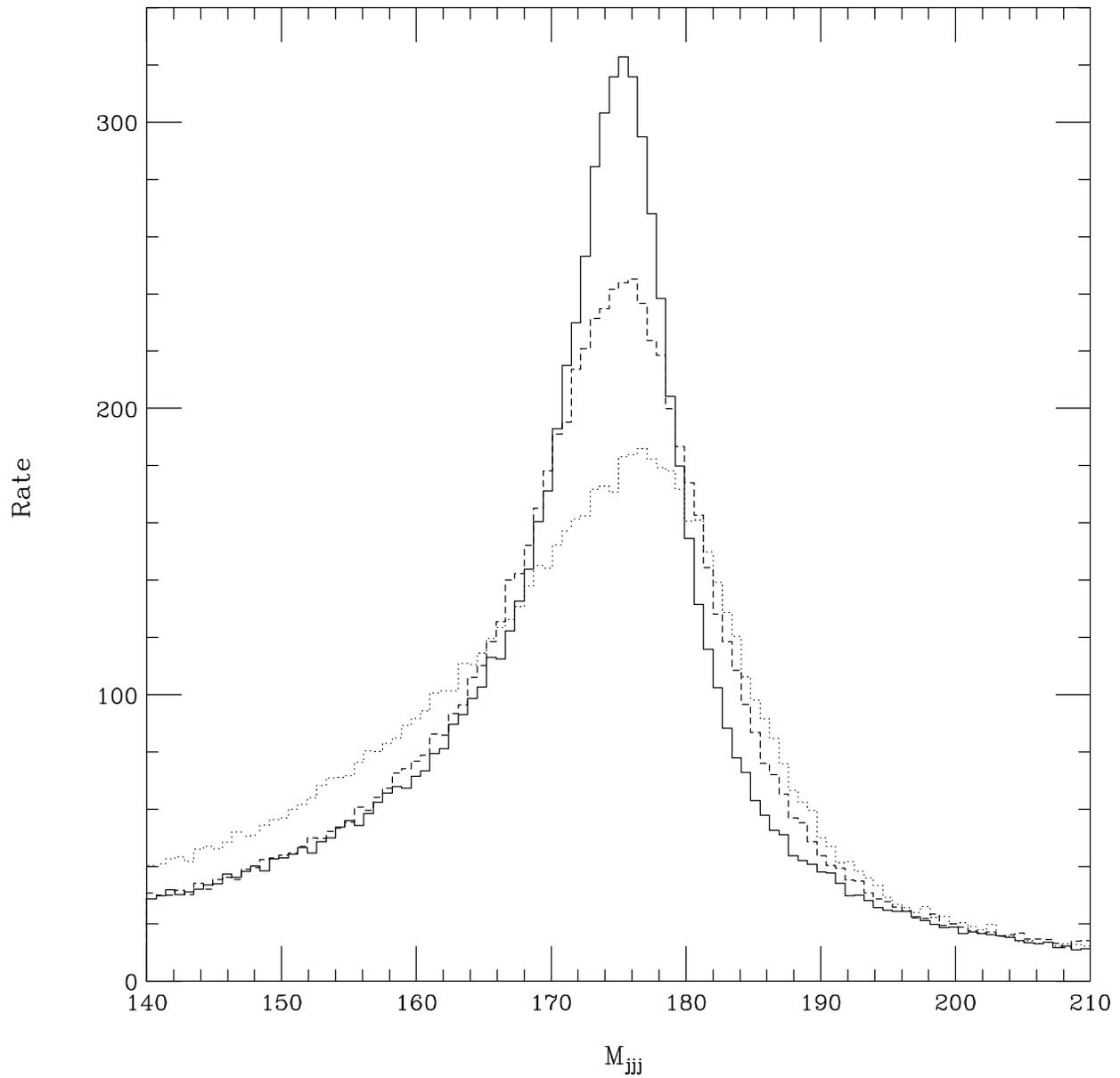}
    \end{center}
\caption{\protect
Similar to Fig.~8, but $M_{\rm jjj}$ is obtained by averaging the
conventional invariant mass and the ``jet angle'' mass measure
of Ref.~\protect\cite{improving}.
}
\label{figure9}
\end{figure}

\begin{figure}
    \begin{center}
       \leavevmode
       \epsfxsize=1.0\hsize
      \epsfbox{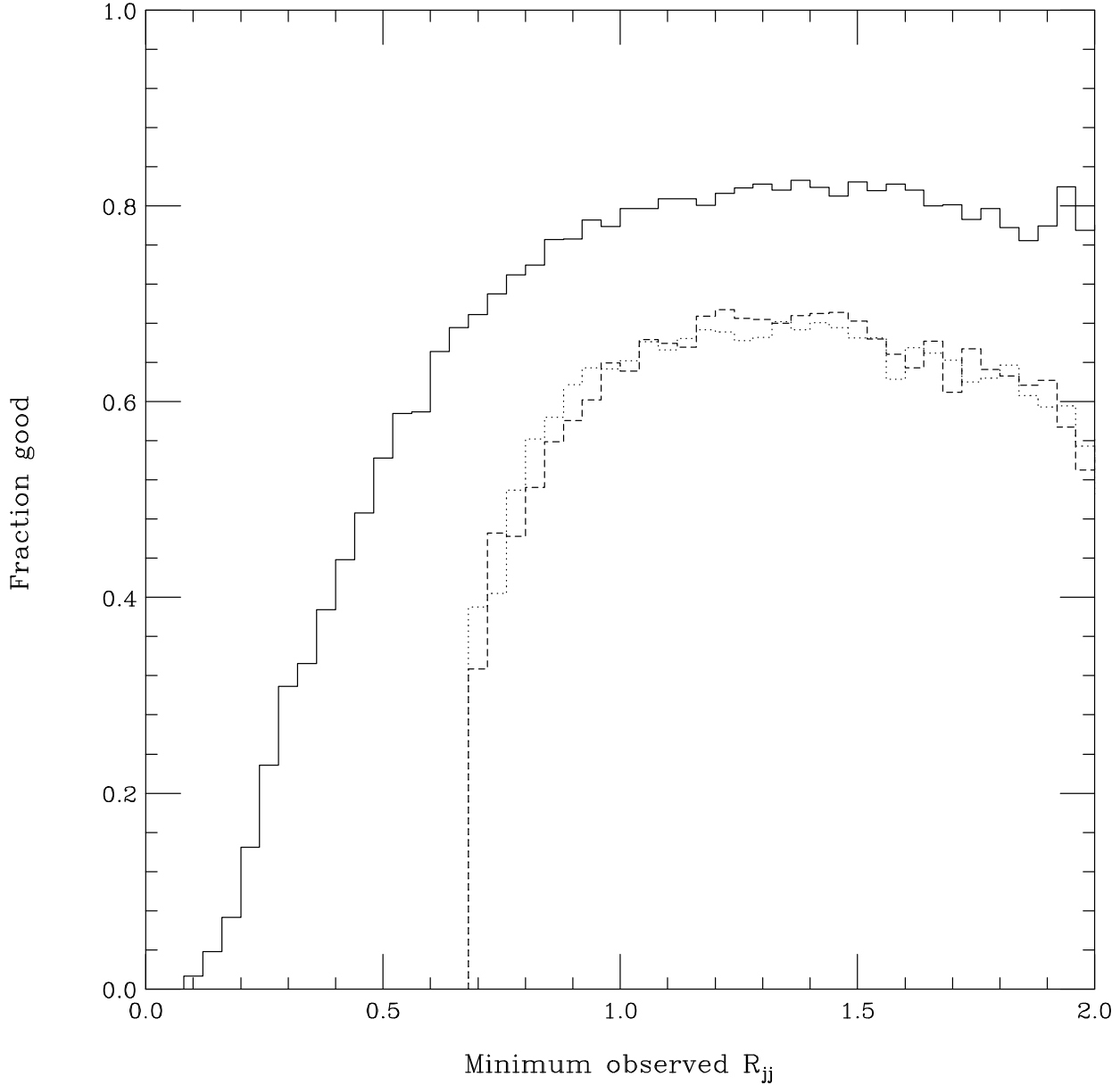}
    \end{center}
\caption{\protect
Fraction of events for which a good match is found between the
4 highest $p_\perp$ observed jets and the 4 primary partons according
to a criterion based on agreement in both angle and energy, as
a function of the minimum separation in $(\eta,\phi)$ between
pairs of observed jets.  Solid curve is for the $k_\perp$ algorithm.
Dashed and dotted curves are for the two versions of cone algorithm
(\protect\cite{howtotell}, dotted \protect\cite{seymour}).
}
\label{figure10}
\end{figure}

\begin{figure}
    \begin{center}
       \leavevmode
       \epsfxsize=1.0\hsize
      \epsfbox{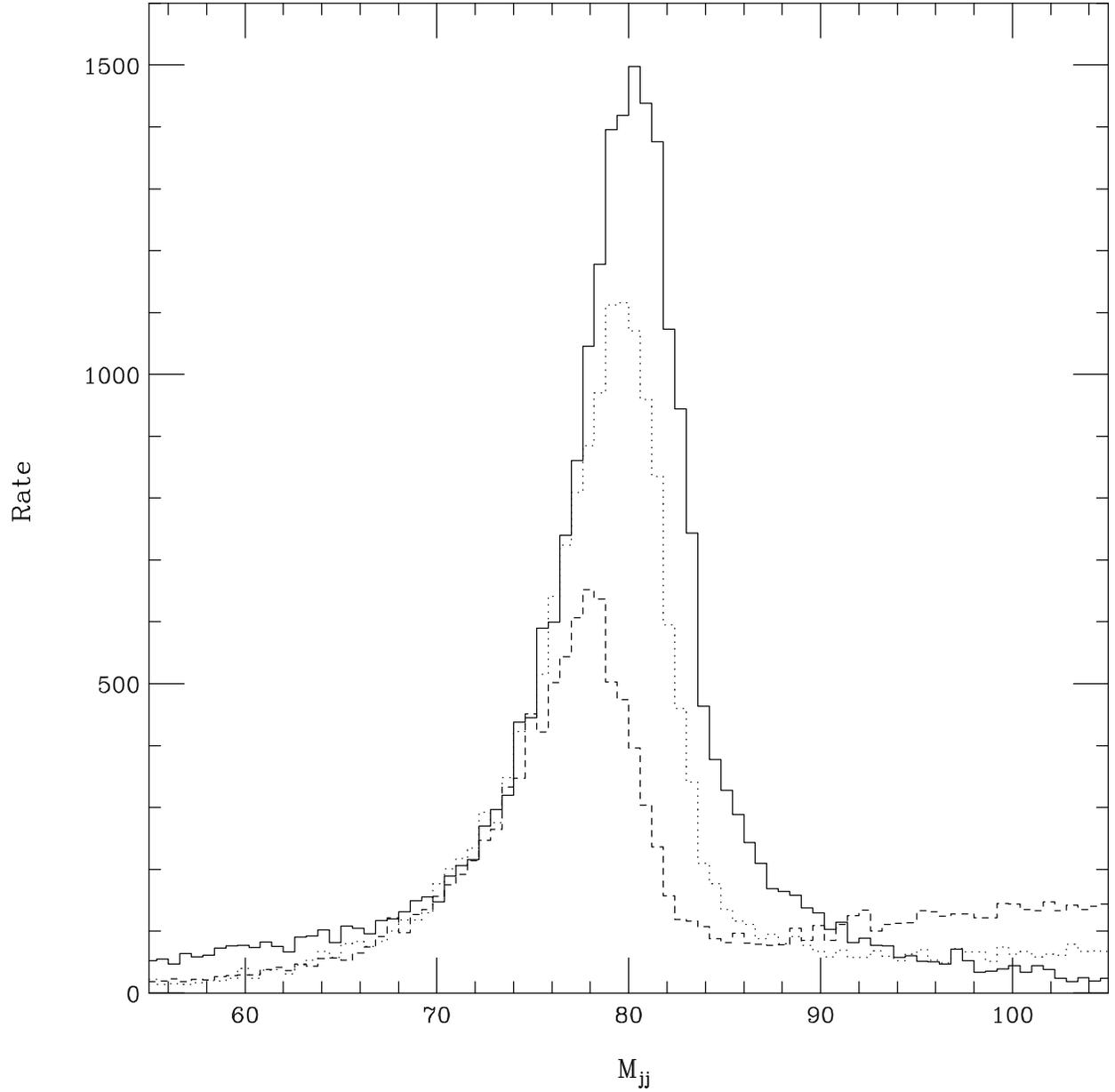}
    \end{center}
\caption{\protect
Dijet mass distribution from $W$ decays identified by
$k_\perp$ algorithm (solid) or cone algorithms
(dashed \protect\cite{howtotell}, dotted \protect\cite{seymour}).
}
\label{figure11}
\end{figure}

\begin{figure}
    \begin{center}
       \leavevmode
       \epsfxsize=1.0\hsize
      \epsfbox{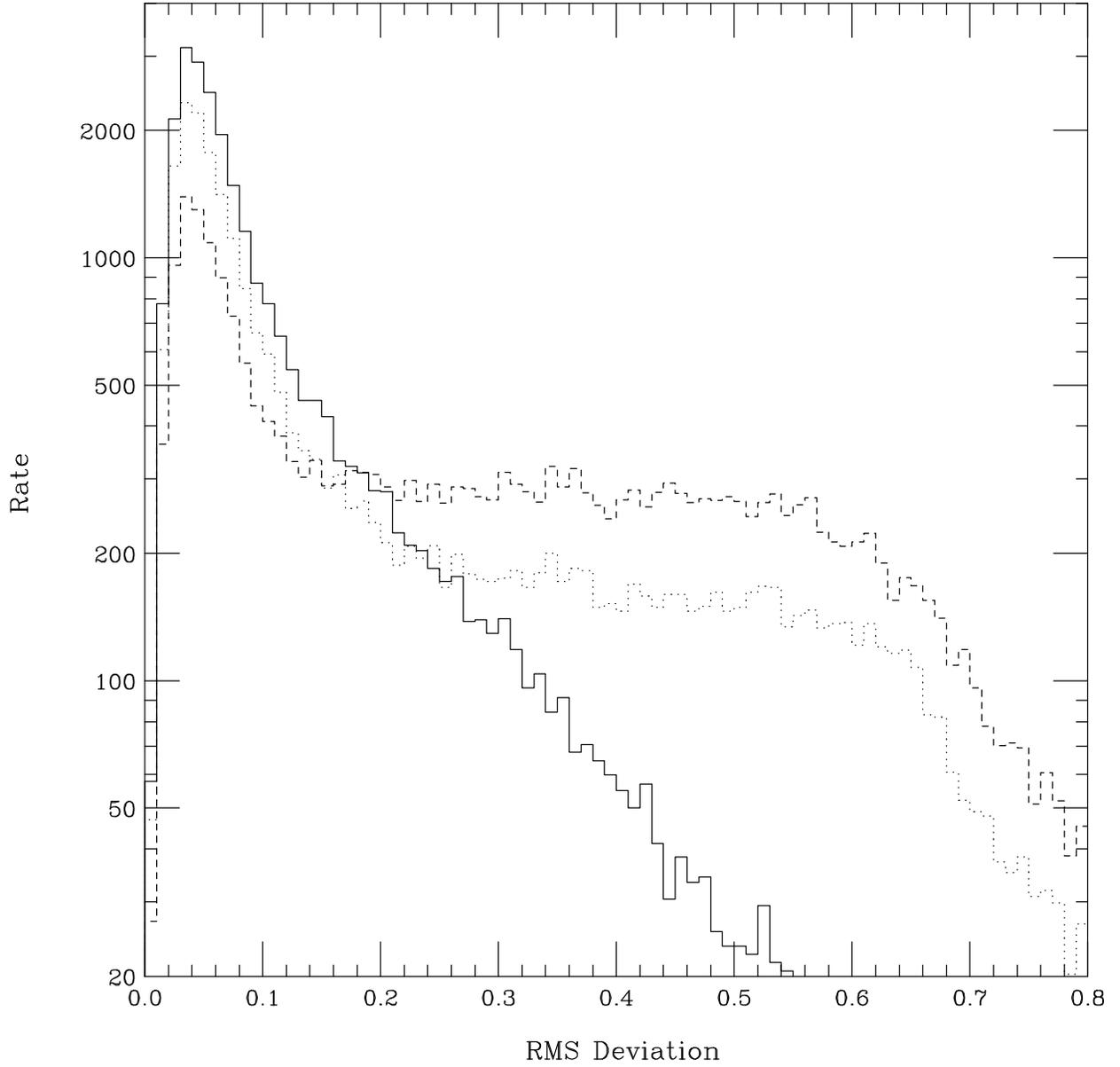}
    \end{center}
\caption{\protect
Distribution of total rms deviation in $(\eta,\phi)$ of best-fitting
dijet pair to $W$ decay partons using
$k_\perp$ algorithm (solid) or cone algorithms
(dashed \protect\cite{howtotell}, dotted \protect\cite{seymour})
as in Fig.~11.
}
\label{figure12}
\end{figure}

\begin{figure}
    \begin{center}
       \leavevmode
       \epsfxsize=1.0\hsize
      \epsfbox{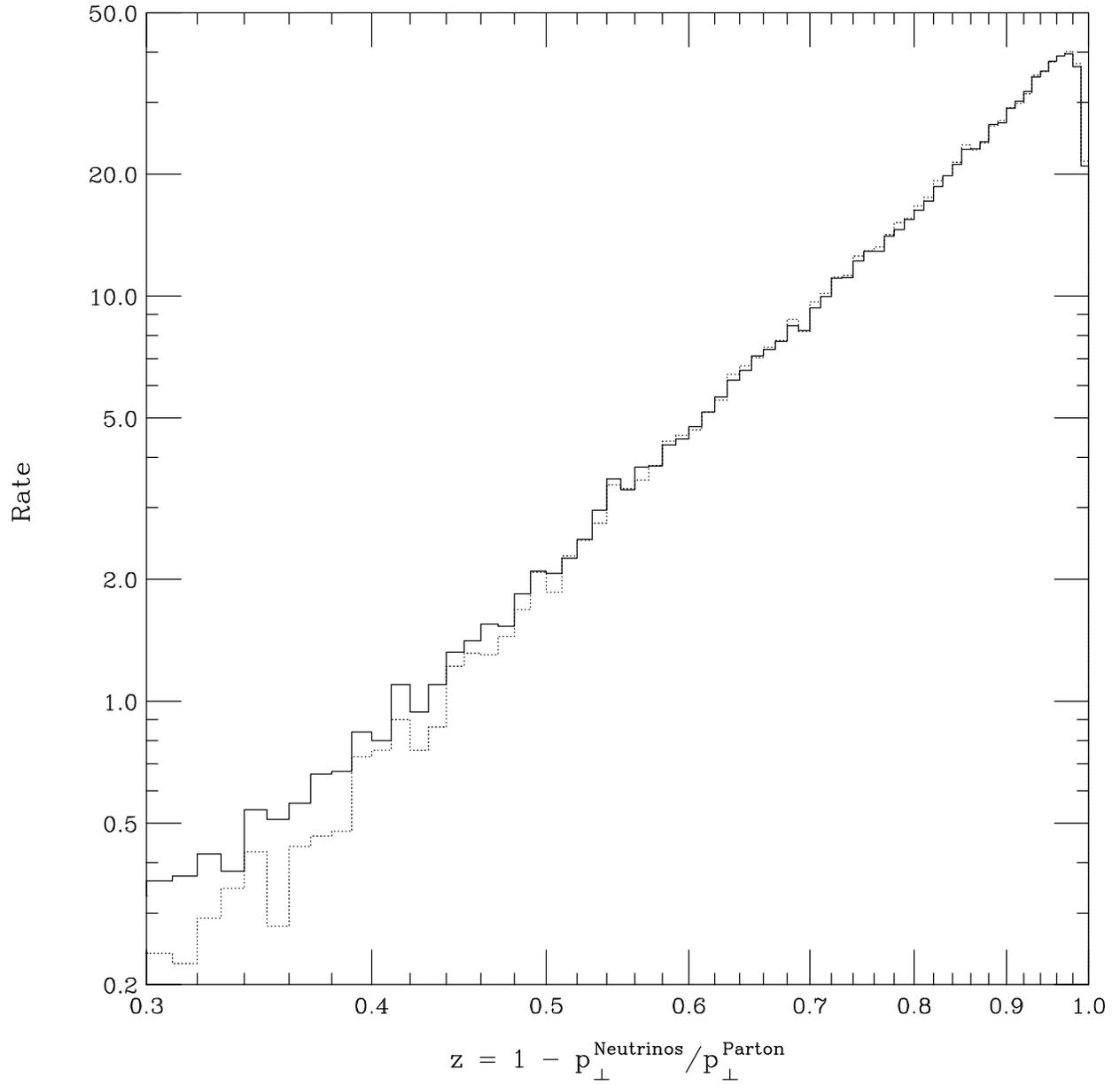}
    \end{center}
\caption{\protect
The observable (i.e., non-neutrino) fraction of the
jet momentum for $b$ jets that contain at least one
neutrino:  solid = all, dotted = jets containing $e^\pm$ or
$\mu^\pm$ of $p_\perp > 2 \, {\rm GeV/c}$.
}
\label{figure13}
\end{figure}

\begin{figure}
    \begin{center}
       \leavevmode
       \epsfxsize=1.0\hsize
      \epsfbox{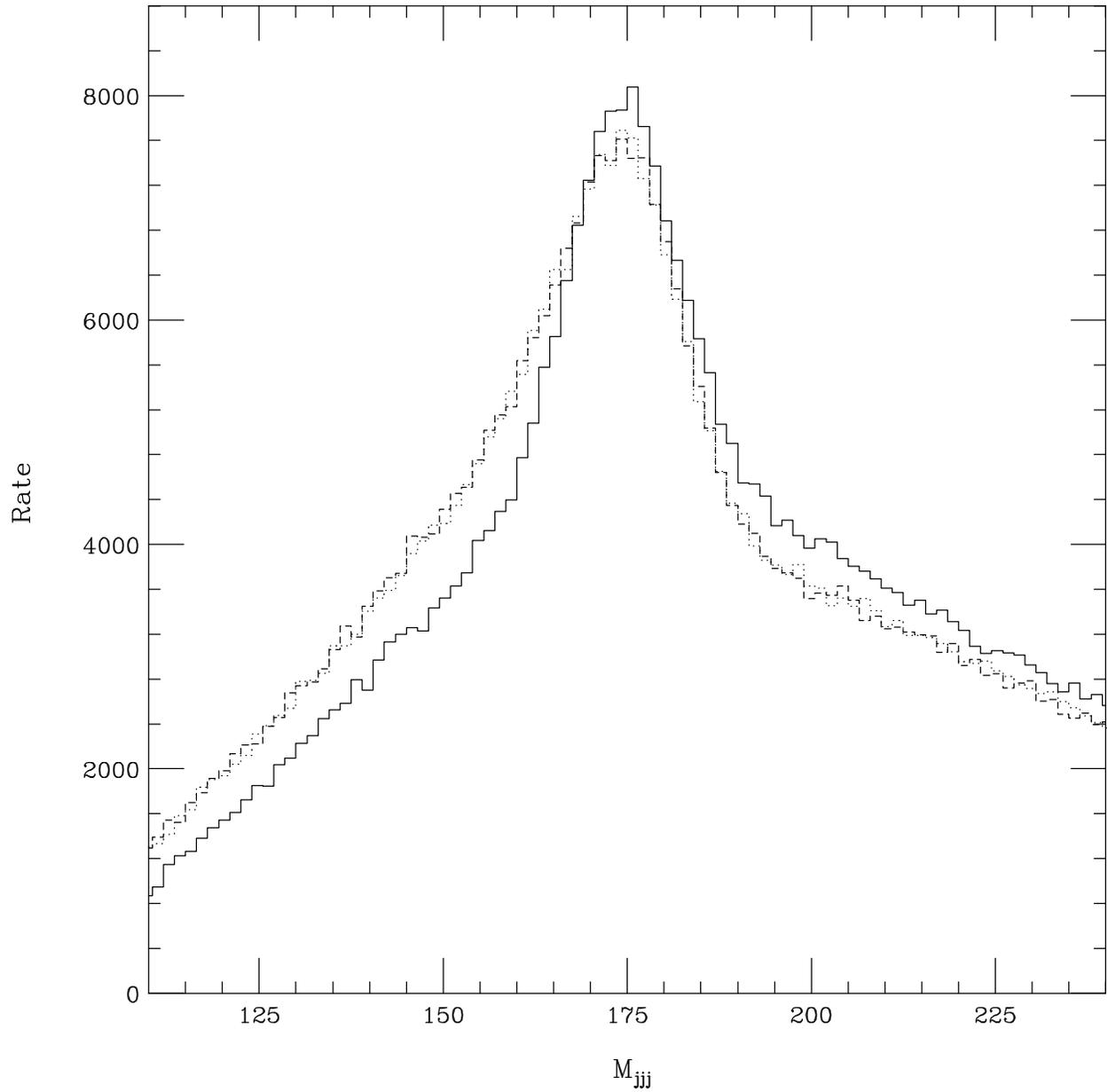}
    \end{center}
\caption{\protect
Trijet mass distributions formed from each 3 of the 4 highest
$p_\perp$ jets observed in each event (4 combinations per event),
using the $k_\perp$ algorithm (solid) or cone algorithms
(dashed \protect\cite{howtotell}, dotted \protect\cite{seymour}).
}
\label{figure14}
\end{figure}

\end{document}